# Spatially conditioned dynamics between population and built form

Anna Brázdová[1] & Martin Fleischmann[1]

[1]Charles University, Department of Social Geography and Regional Development

**Abstract**

Understanding the relationship between population and the built environment is essential for addressing socio-spatial inequalities. While researchers have long theorized these dynamics, empirical analyses remain limited. This study proposes a spatially explicit framework to quantify the relationship between population and the built environment at the scale of local census tracts in Czechia. The approach integrates a fine-grained classification of built form with a comprehensive set of socio-demographic indicators. The method compares global and geographically weighted classification models to assess the overall strength and spatial variability of the associations between population structure and built form. The results of the study show that population characteristics exhibit linear, spatially conditioned relationships with built form, emphasizing that spatial heterogeneity must be accounted for when assessing these relationships. The analysis also reveals that some built form types are more socially selective than others, underscoring the importance of built form in reproducing social-spatial inequalities.

# 1 Introduction

The physical layout of buildings, streets, and plots affects how people move around, how they interact, how they access opportunities, how built-up space is perceived, and how different social groups are distributed across space (Oliveira, 2016). The place where people live and its built environment is an active component in socio-spatial differentiation, where morphological configuration not only reflects existing social structures but also actively creates the conditions that reinforce social stratification (Harvey, 1973; Soja, 1980; Talen, 2008). In this context, urban form, or built form—inclusive of non-urban environments—provides spatial affordances – different environmental conditions create certain opportunities and possibilities, while placing constraints on others (Gibson, 1979). This perspective raises the question of whether different types of built form exhibit different capacities to host diverse socio-demographic compositions.

Understanding how people and the built environment interact has long been a subject of research in urban studies. Traditional research has often focused on qualitative case studies based on observation or personal views (Jacobs, 1961; Whyte, 1988), which provide only a limited view of the relationship. The primary limitation of qualitative methodologies lies in the difficulty of replicating and generalising the findings. More recently, researchers adopted quantitative methods to study the relationship between features of the urban environment and socio-economic aspects, but remain constrained by methodological limitations—the main one being the difficulty of representing the built form itself.

Capturing the complex nature of built form at scale has historically been labor-intensive and methodologically challenging. Built form has been studied in many ways, from the traditional, primarily qualitative, schools of urban morphology, to quantitative and computational approaches that operationalise the built environment through measurable morphological attributes, spatial metrics, and data-driven classifications (Oliveira, 2016). The first significant advancement in systematic, quantitative approaches came with the application of the Space Syntax Laboratory in the mid-1980s, which enabled scalable numerical description of street networks and revealed connections between network structure, land-use diversity, and socio-economic conditions (Hillier & Hanson, 1984; Vaughan et al., 2005). However, this focus on a single component of built form limits the understanding of its broader complexity.

Moreover, the analysis of the interplay between the built environment and socio-economic dimensions of human settlements is more manageable when the built form is represented by a small number of interpretable types. Significant progress in this direction has been made thanks to the open data revolution (Kitchin, 2014) and the emerging field of urban morphometrics – a set of methods aimed at characterising built form based on its measurable traits (Fleischmann et al., 2025). These methods produce detailed typologies that differentiate between different historical layers, distinguishing medieval from modernist urban developments, single-family housing from row houses, and similar (Araldi & Fusco, 2025; Fleischmann et al., 2022). Starting from the scale of neighbourhoods, these methods evolved into scalable analytics that characterise entire countries while retaining the granularity of individual features (Araldi & Fusco, 2025; Fleischmann et al., 2022).

Despite these advances, the research direction that can capture both socio-economic nuances in the population and urban morphology in its complexity has started to appear only recently, largely through integrated, multidisciplinary frameworks. Studies of Venerandi et al. (2024) quantifying the relationship between urban form and deprivation for the city of Isfahan; Sapena et al. (2021) linking quality of life to morphology (albeit using a simplified typology of Local Climate Zones (Stewart & Oke, 2012)); Duque et al. (2015) linking crime and intra-urban poverty to remotely sensed landscape metrics; or Oliveira (2024) examining the correlation between urban form and socio-economic and environmental diversity, collectively demonstrate that urban form interacts with social dimensions.

However, most of these studies rely on simplified representations of urban form and are typically confined to relatively small geographical extents—often individual neighbourhoods or cities—and focus on isolated phenomena such as quality of life or deprivation, constraining their capacity to systematically quantify the associations between population and built form at scale. Moreover, existing work largely assumes spatial stationarity in the relationship between built form and population, implying that this relationship remains constant across the entire study area. Such assumptions are problematic in geographical research, as social processes are deeply embedded in space and vary across neighbourhoods, cities, and regions due to historical trajectories, institutional arrangements, and local urban morphologies (Harvey, 1973; Soja, 1980).

The literature shows that urban form exhibits consistent, quantifiable relationships with socio-economic conditions, demonstrating the potential of integrated analyses, while also highlighting the need for more comprehensive frameworks. The prevalent reliance on global statistical models, individual case studies, and simplified typologies limits the generalisability, scalability, and transferability of findings. Therefore, a research framework is needed to move beyond these issues and incorporate local, multi-level techniques capable of adequately capturing the complex relationship between built form patterns and the populations inhabiting them. Without such a change, our understanding of the relationship will remain incomplete.

This study quantitatively assesses the strength and spatial variability of associations between population and built form in order to capture spatially conditioned dynamics between the two. It develops a methodological framework that compares global and geographically weighted classification models to assess the associations between built form and population. It uses the Urban Taxonomy classification (Fleischmann et al., 2025) derived from urban morphometrics as a proxy for the physical environment, along with a comprehensive set of socio-demographic indicators derived from the decennial census to represent population structure. Specifically, the study addresses the following research questions: i) How closely is morphology tied to population structure? ii) How do the associations differ across different built form types? iii) Are the associations consistent or spatially variable?

The contributions of this study are both methodological and empirical. Methodologically, it offers a framework composed of semi-automated variable selection combining factor analysis and a minimum spanning tree, and geographically weighted classification modelling for studying built form–population associations. Empirically, it quantifies the strength and spatial variability of these associations across Czechia, providing insight into how different built forms and geographic settings interact with socio-demographic structure in a post-socialist context.

The remainder of this paper is structured as follows. The next section details the data and methodology employed. Afterwards, we present the empirical results, followed by a discussion of key findings and a conclusion in the last section.

# 2 Methods

This study adopts a multi-step analytical approach (outlined in Figure 1) to investigate the associations between built form and population characteristics at the scale of local census tracts (named Basic Settlement Units) in Czechia. First, we outline a semi-automated variable selection method to identify the most relevant variables. Second, we assess the associations through an analytical framework, which we begin by developing baseline predictive models. We then examine the spatial autocorrelation of models' errors to detect any unaccounted-for spatial patterns. Finally, we develop geographically weighted models and evaluate their performance relative to the baseline, providing insight into the strength and spatial variability of the associations between built form and population characteristics.

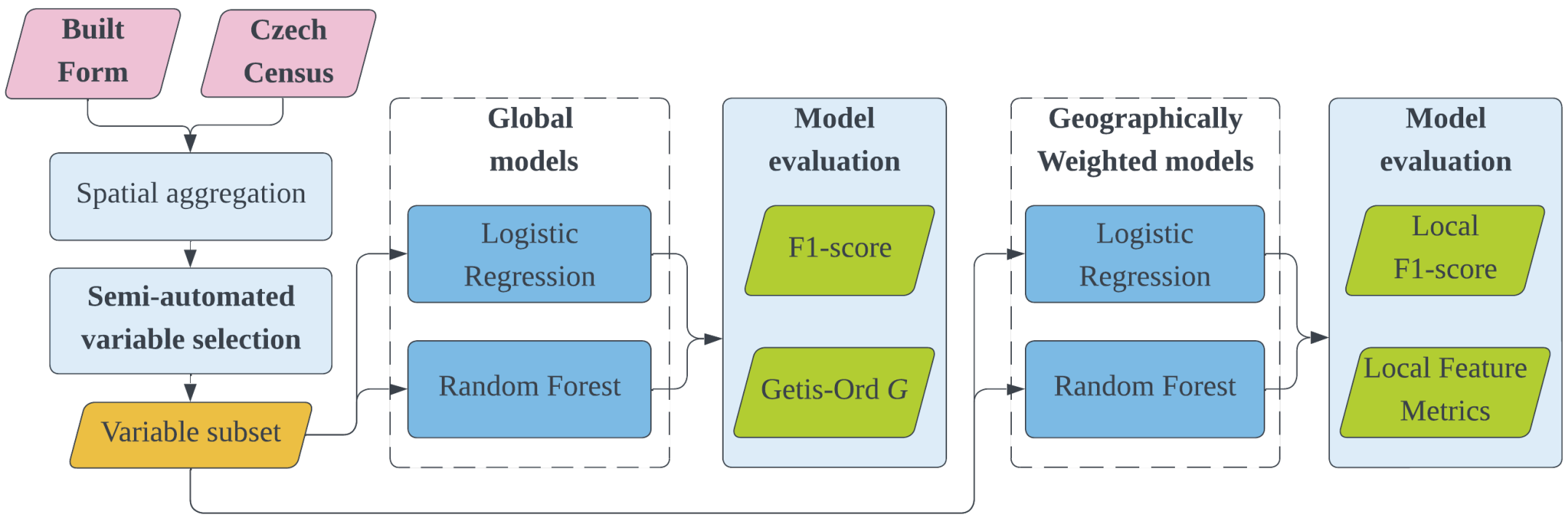

*Figure 1: Methodological workflow of the analytical framework. The workflow illustrates the integration of two data sources, data preprocessing, and comparative application and evaluation of global and geographically weighted models.*

## 2.1 Data

The work relies on two data sources: a classification of built form and socioeconomic data from the Czech National Census.

The built form categories are derived from Urban Taxonomy (Fleischmann et al., 2025), an open morphometric classification in which each building belongs to a distinct category, with each category representing recurring patterns of built form. The data product is a specific outcome of the Hierarchical Morphotope Classification method (Fleischmann et al., 2025), which privides a scalable, systematic, and fine-grained typology derived from the concept of the morphotope – the smallest locality with a distinctive character (Conzen, 1988). By capturing the combined characteristics of buildings, plots, streets, and their spatial configuration, the approach organises built form into a hierarchical taxonomy, enabling classifications to be derived at different levels suited to multiple analytical scales, from detailed local typologies to more general regional patterns. For this study, we use Level 3 of the taxonomy, which includes 8 categories (shown in Figure 2) that capture a variety of built forms at the national level.

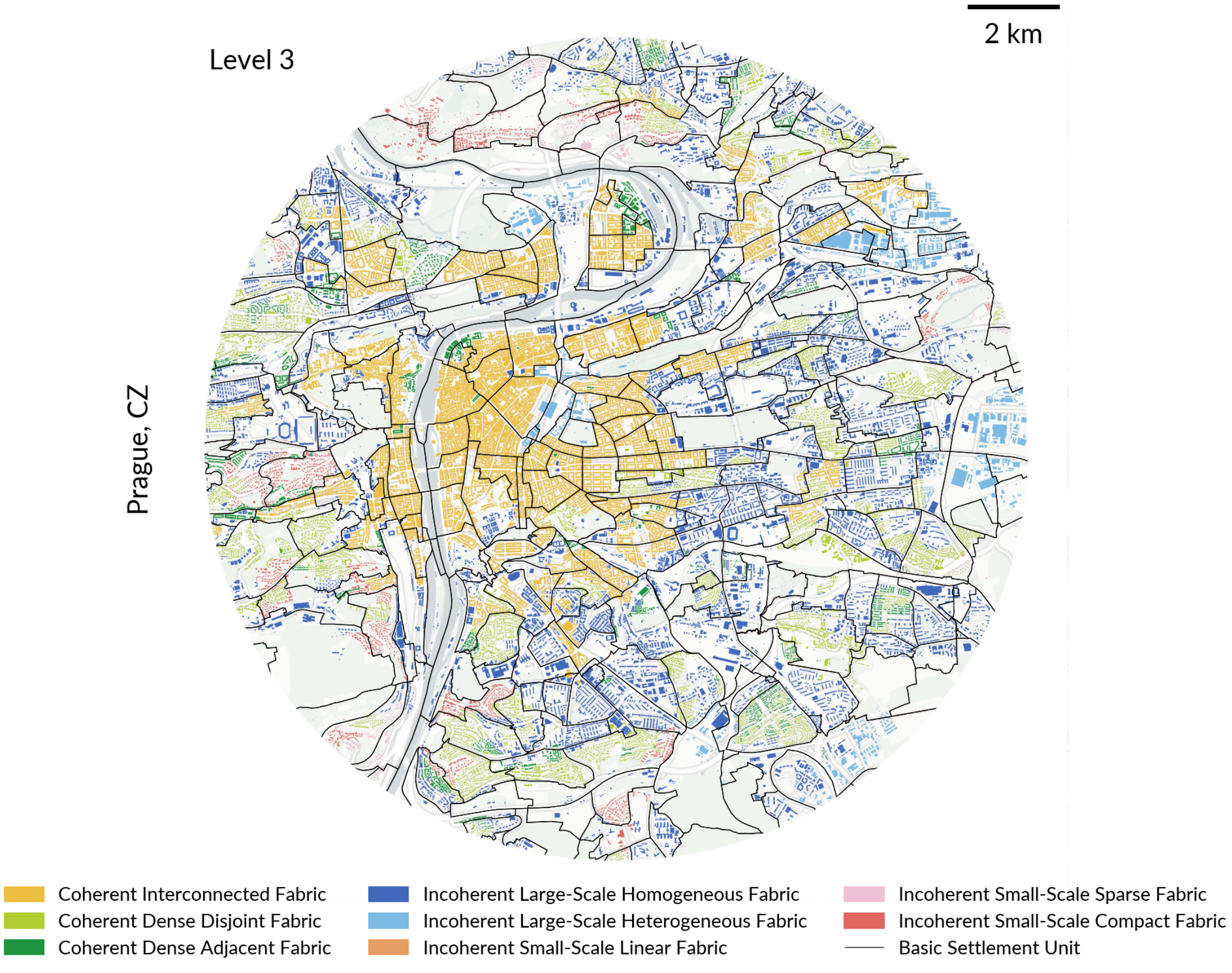

*Figure 2: Level 3 taxonomy with BSU boundaries showing the morphological homogeneity within these areas, illustrated in Prague, CZ*

The social structure is characterized by the latest Czech National Census (CNC) (ČSÚ, 2021), providing detailed information about the economic, social, and demographic characteristics of the population. The socio-demographic variables are collected at the scale of basic settlement units (BSU) – the smallest statistical units available to researchers. With the median population of approx. 150 inhabitants, these units allow for a very granular spatial resolution.

Given that Urban Taxonomy is defined at the building level, and data related to population are available at the level of Basic Settlement Units, a spatial aggregation step is required. BSUs represent locations with clear territorial, technical, and urban characteristics, meaning that their boundaries historically reflect underlying settlement structures (ČSÚ, 2014). The predominant types of built form within each unit usually correspond to its dominant built-up structure and therefore provide a suitable approximation of the local morphological context. Therefore, we aggregate building-level classifications into BSUs by assigning the most common type of built form.

## 2.2 Variable selection using Factor Analysis and Minimum Spanning Tree

High-dimensional census data (CNC 2021 provides 788 variables) often contain correlations between variables, overlapping measurements of similar phenomena, or indicators that provide little unique information. Including such variables can reduce model's reliability and interpretability, negatively affecting performance due to the curse of dimensionality (Guyon & Elisseeff, 2003; Rojas, 2015). Including a larger number of variables enhances predictive or explanatory power only when those variables are reliable, robust, and provide new, non-redundant information. Although dimensionality reduction techniques can compress data into a small number of underlying components that account for a significant proportion of variance, these components represent different combinations of the original data and provide only limited insight into identifying the important variables that influence the observed patterns (Harris et al., 2005). On the other hand, the objective of variable selection is to identify the smallest subset of input variables that captures the greatest variability in the original dataset (Gale et al., 2016; Harris et al., 2005). This process usually requires balancing theoretical and empirical considerations. The selection should be based on a broader literature review to account for potential influences on area differentiation. From an empirical perspective, it is necessary to evaluate individual indicators in terms of redundancy, correlation structure, distribution properties, and spatial coverage (Spielman & Singleton, 2015).

Our framework builds on the proposed method for Principal Component Analysis (PCA)-based automated variable selection of Liu et al. (2019); however, instead of PCA, we use Factor Analysis (FA), which has been widely used to identify the underlying dimensions of socio-demographic structure (Clark et al., 1974; Janson, 1980). Whereas PCA prioritizes explaining the total variance of the original variables, FA distinguishes between common and unique variance, attempting to explain the covariance among the original variables and accounting for the variance shared between variables (Schreiber, 2021). Given the nature of the population data, we consider socio-demographic characteristics to be manifestations of underlying social structures, such as *socio-economic status* or *household lifecycle*, rather than independent numerical observations. In this context, PCA risks omitting variables with lower total variance but high structural importance. Factor analysis is more suitable as it allows identification of variables that contribute most significantly to the latent structures of the population. The framework comprises multiple stages, which are illustrated in Figure 3.

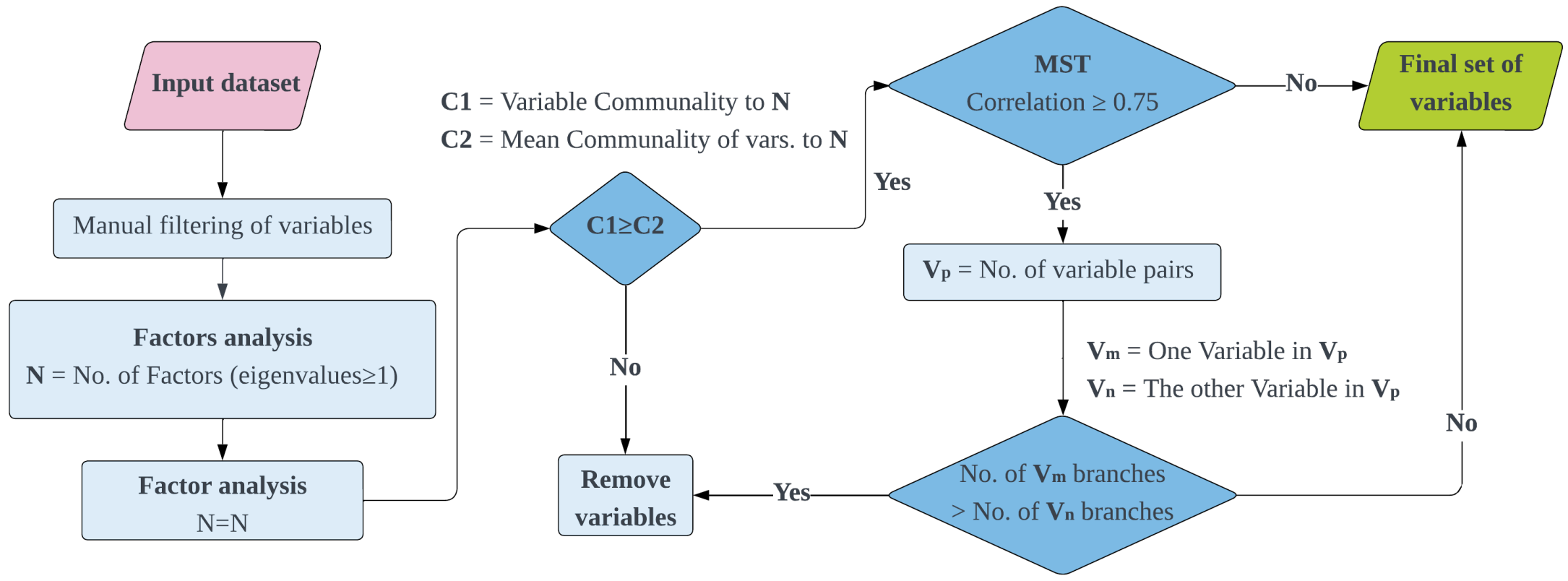

*Figure 3: Workflow of the variable selection method using Factor Analysis and Minimum Spanning Tree.*

First, we filter variables based on theoretical and practical relevance, excluding overlapping measurements, measurements of similar phenomena, or measurements linked to buildings and housing conditions due to potential conceptual overlap with built form classification.

In the second stage, we generate a set of latent factors from the input variables. The number of factors to be retained ($N$) is determined using Kaiser's criterion, selecting factors with eigenvalues greater than or equal to 1 (Kaiser, 1960). The contribution of each variable to the retained factors is quantified using communalities ($C_1$), which represent the proportion of its common variance explained by the $N$ retained factors. The communality value ranges from 0 to 1, with larger values indicating a stronger influence of the variable on the factor. The mean communality ($C_2$) is calculated as a reference point. Following the methodology of Liu et al. (2019), variables with communality below $C_2$ are removed to ensure only those with a significant contribution to the latent factors are retained.

The final stage explores the correlation among the retained variables using a Minimum Spanning Tree (MST), which re-examines the level of data redundancy. The MST is constructed from the variable correlation matrix, where nodes represent individual variables and edge weights represent dissimilarity. The tree highlights pairs of variables ($V_p$) that remain highly correlated, using a correlation coefficient greater than or equal to ±0.75, which is commonly cited as the "rule of thumb" for high correlation (Udovicic et al., 2007). For each identified pair, the MST assesses the connectivity of the two nodes ($V_m$ and $V_n$) within the network, and variables with lower connectivity are removed from the list of candidate variables, as they are considered less informative (Liu et al., 2019).

To evaluate the robustness of the variable selection framework, we benchmarked the global classification models described in the section 2.3, using both the original dataset containing all socio-demographic variables, and the subset derived from the variable selection process. The results showed that the subset achieved performance comparable to that of the full dataset, suggesting that the variable selection framework successfully reduced dimensionality and eliminated redundancy without compromising discriminatory power.

## 2.3 Association between built-form and population characteristics

We assess the association between built form and population characteristics through an analytical framework comprising five sequential steps. First, we train a set of global classification models as a baseline. Second, we evaluate the spatial distribution of the models' errors to determine whether they exhibit spatially homogeneous or heterogeneous patterns. Third, we train geographically weighted classification models—one for each built-form type—to capture the spatial variability of these relationships and examine their performance relative to the baseline. This assesses the extent to which the population structure is associated with the built environment and the degree of spatial heterogeneity in the relationships. Finally, we explore local feature metrics of the geographically weighted models to identify variables that consistently have high discriminatory power and help distinguish between built-form types.

For the baseline models, we compare Logistic Regression (LR) and a Random Forest (RF) classifier to test whether the relationship under scrutiny follows linear or non-linear behaviour. To mitigate the effects of spatial autocorrelation and uneven distribution of built form types, both models were evaluated through spatial block cross-validation (Roberts et al., 2017), where BSUs were grouped by their administrative county (LAU1) boundaries to ensure geographic separation of training and test sets.

Next, we compute the Getis–Ord $G$ statistics (Getis & Ord, 1992) on the errors of the global models to evaluate whether spatial autocorrelation affects these global estimates. The Global $G$ Autocorrelation Statistic is used to assess the overall clustering in the dataset, and the Local $G$ Autocorrelation Statistic is used to detect patterns of spatial autocorrelation that are not evident with global statistics (Ord & Getis, 1995). It enables us to identify areas where errors are concentrated.

To accommodate spatial heterogeneity, we propose the use of Geographically Weighted Classification, a type of spatial model that takes non-stationary variables into consideration and models the local relationships between predictors and an outcome of interest. We rely on the agnostic framework provided by the Python package *gwlearn* (Fleischmann et al., 2026), which identifies a local subset of training data around each geometry and fits a local estimator minimising a weighted empirical loss, where the weights depend on spatial distance. Mirroring the baseline models, we use LR and RF as local estimators. The result is an ensemble of local models, which allows local coefficients or decision trees within GWRF to vary across space, naturally reflecting spatial non-stationarity. Given that we are dealing with a classification task, rather than a regression predicting continuous variables, the framework comes with a set of limitations. The most notable one is the restriction of such a model to binary classification, rather than multi-class classification as in global models. This requires fitting 8 independent models, one for each category, for each model architecture (GWLR, GWRF), where each model predicts whether a given BSU belongs to a class or not. For each, we empirically determine the optimal bandwidth used to define the local neighbourhood.

To evaluate the performance of the global models we use the macro-averaged F1 score which balances precision and recall to provide a single measure of model performance. While accuracy can be inflated by high performance in dominant, more common classes, F1-macro weights all classes equally and is therefore less sensitive to class imbalance. For geographically weighted models, the macro-averaged F1 score was computed for each local model and averaged to produce a global performance estimate.

Finally, the local coefficients and local feature importances of the geographically weighted models are used to interpret the classification outcomes. This allows us to identify which socio-demographic variables carry the strongest influence in distinguishing between built form types, and whether this influence is consistent across space or varies geographically. Furthermore, it enables us to assess which built form types exhibit the greatest sensitivity to spatial context.

# 3 Results

In this section, we first present the results from the automated variable selection framework. Second, we illustrate the outcomes of the classification analyses. Third, we offer interpretations for the behaviours of the local feature metrics.

## 3.1 Variable selection using Factor Analysis and Minimum Spanning Tree

The variable selection process presented in the previous section was applied to the Czech census data. As the original census datasets contained a total of 788 variables, we first filtered out variables based on theoretical and practical relevance. In practice, all variables related to buildings and their conditions were removed, alongside duplicated or similar variables caused by various partitioning options or embedded measurements of the same phenomena. This filtering yielded 89 variables, organised into three domains: demographics, socio-economic, and housing.

Factor Analysis of the input data resulted in 26 meaningful factors, of which 38 variables had communalities greater than the overall average. Of the 38 variables initially identified, 10 pairs were highly correlated (correlation coefficient $\geq \pm 0.75$); therefore, using the MST, 10 variables were removed. This resulted in a final set of 28 variables, listed in Table 1. This subset serves as a representative sample of the initial dataset, capturing relevant information across all three domains.

*Table 1: Retained variables for modelling.*

| Demographics | Socio-economic | Housing |
|---|---|---|
| Age Group: 0-6 | Education - Secondary (without graduation) | Residents in Owner-Occupied Dwellings |
| Age Group: 7-14 | Education - Secondary (with graduation) | Residents in Rented Dwelling |
| Age Group: 15-24 | Education - University | Residents in Cooperative Dwellings |
| Age Group: 45-54 | Unknown Education | Residents in Privately Owned Houses |
| Citizenship - Slovakia | Empl. sector - Agriculture/Forestry/Fishery | Persons per Dwelling |
| Citizenship - EU | Empl. sector - Industry | |
| Citizenship - Unknown | Empl. occupation - Service/Sales | |
| Long-Term Stay Residents | Empl. occupation - Craft/Repair | |
| Religion - Non-Religious | Empl. status - Employees | |
| Religion - Unspecified | Econ. Activity - Parental Leave | |
| Marital Status - Married | | |
| Marital Status - Divorced | | |
| Marital Status - Widowed | | |

## 3.2 Association between built-form and population characteristics

### 3.2.1 Global models

To assess the extent to which population and built form are associated, we first trained a set of global classification models. Two global modelling strategies were tested: a linear model and a non-linear model, and their performance was evaluated using the macro-averaged F1 score. The higher the score, the more the spatial distribution of the population structure is linked to the spatial distribution of the built form. Initial results (Table 2) indicated that the associations are difficult to capture at the global level. The LR and RF models yielded low macro F1 scores of 0.26 and 0.33, respectively, with class-level F1 scores varying considerably across built form types. Although the *Incoherent Small-Scale Sparse Fabric* class achieved a relatively strong F1 score of 0.61, performance across most other classes remained low to moderate (0.33–0.47), dropping as low as 0.11–0.19 for two specific categories.

### 3.2.2 Spatial autocorrelation

We used the Getis–Ord $G$ statistic on the errors of the global models to evaluate whether there is significant spatial non-stationarity that these global models are unable to account for. The Global $G$ test showed statistical evidence of spatial autocorrelation in the models ($p$ = 0.001). As illustrated in Figure 4, the contiguous clusters of statistically significant local autocorrelation of errors are concentrated in specific geographic regions across the country. Both LR and RF models exhibit these clusters in the suburban fringes surrounding the two metropolitan centers - Prague and Brno, the peripheral border regions of Bohemia, and across eastern Moravia and Silesia.

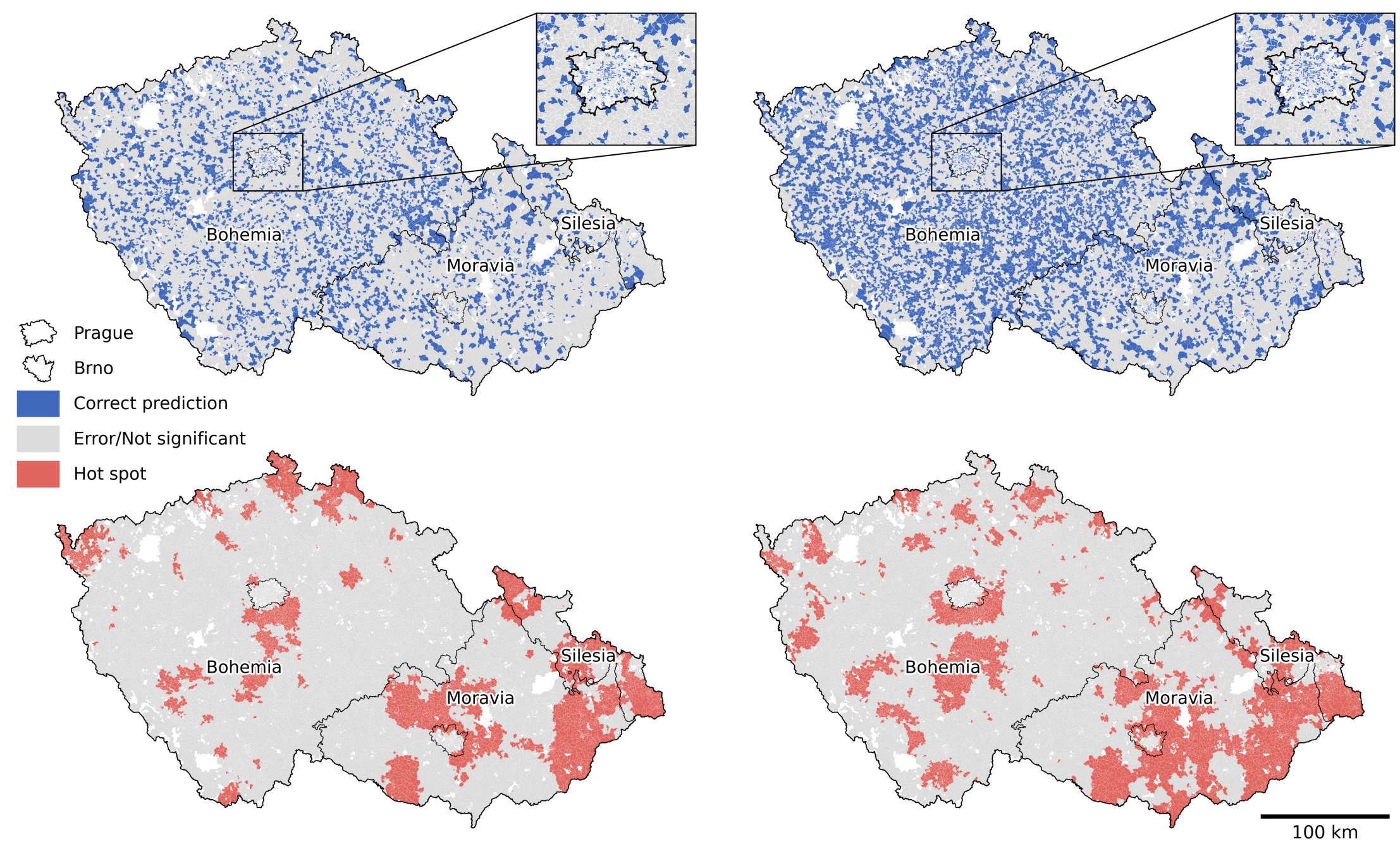

*Figure 4: Comparison of predictive performance and spatial patterns of error between Logistic Regression (left) and Random Forest (right) models. The top maps illustrate classification results, with insets highlighting the Prague metropolitan area. The bottom maps display the spatial autocorrelation of the errors, identifying statistically significant hot spots.*

Because of the limited performance of global models and the statistically significant spatial autocorrelation of errors, global modelling was considered insufficient for capturing the spatially variable relationship between population characteristics and built form types.

### 3.2.3 Geographically weighted classification

Given the presence of spatial autocorrelation in the errors, we trained a series of geographically weighted classification models that allow parameters to vary locally, using both linear and non-linear models, with a separate model estimated for each built-form type.

All geographically weighted models showed moderate to high F1-macro scores, as shown in Table 2.

*Table 2: Averaged F1-macro score of geographically weighted LR and RF models compared to F1-macro score of global LR and RF model.*

| | ***LR*** | | | ***RF*** | | |
|---|---|---|---|---|---|---|
| | global | GW | *Δ* | global | GW | *Δ* |
| Incoherent Large-Scale Homogeneous Fabric | 0.41 | 0.83 | +0.42 | 0.47 | 0.81 | +0.34 |
| Incoherent Large-Scale Heterogeneous Fabric | 0.12 | 0.76 | +0.64 | 0.14 | 0.67 | +0.53 |
| Incoherent Small-Scale Linear Fabric | 0.28 | 0.68 | +0.40 | 0.19 | 0.63 | +0.44 |
| Incoherent Small-Scale Sparse Fabric | 0.33 | ***0.66*** | +0.33 | ***0.61*** | ***0.66*** | ***+0.05*** |
| Incoherent Small-Scale Compact Fabric | 0.23 | 0.63 | +0.40 | 0.33 | 0.63 | +0.30 |
| Coherent Interconnected Fabric | 0.32 | 0.89 | +0.57 | 0.45 | 0.85 | +0.40 |
| Coherent Dense Disjoint Fabric | 0.27 | 0.72 | +0.45 | 0.33 | 0.74 | +0.41 |
| Coherent Dense Adjacent Fabric | 0.09 | 0.62 | +0.53 | 0.11 | 0.62 | +0.51 |

For most built form types, the performance of the geographically weighted linear and non-linear models is almost identical. In the case of *Incoherent Large-Scale Homogeneous Fabric*, *Incoherent Large-Scale Heterogeneous Fabric*, and *Coherent Interconnected Fabric*, GWLR even outperforms GWRF. The only exception is *Coherent Dense Disjoint Fabric*, where GWRF showed a marginal improvement. Non-linear models do not capture any additional structure, as linear models provide a sufficient level of distinction (with F1-macro scores ranging from approximately 0.62 to 0.89), showing that the associations between population characteristics and built form types are predominantly linear on the local level.Given the low standard deviation (≈5 %) of the F1-macro score across different locations, the associations between population and built form are comparably strong across the study area for all built-form types.

However, for *Incoherent Small-Scale Sparse Fabric*, the global RF achieved similar performance to the GWLR and GWRF models (F1-macro 0.61 and 0.66, respectively), indicating that in this case, even though the relationship is locally linear, it can be captured by a non-linear model at the global level.

### 3.2.4 Local feature metrics

To identify which variables have the strongest overall influence on distinguishing between built form types, we extracted the values of the regression coefficients from the GWLR models and feature importances from the GWRF models. In Figure 5, we present the distributions of these metrics, ordered by the mean absolute value of regression coefficients, to show the magnitude of their effect. Each variable domain (demographics, socio-economic, and housing) is represented with a distinct colour for comparison.

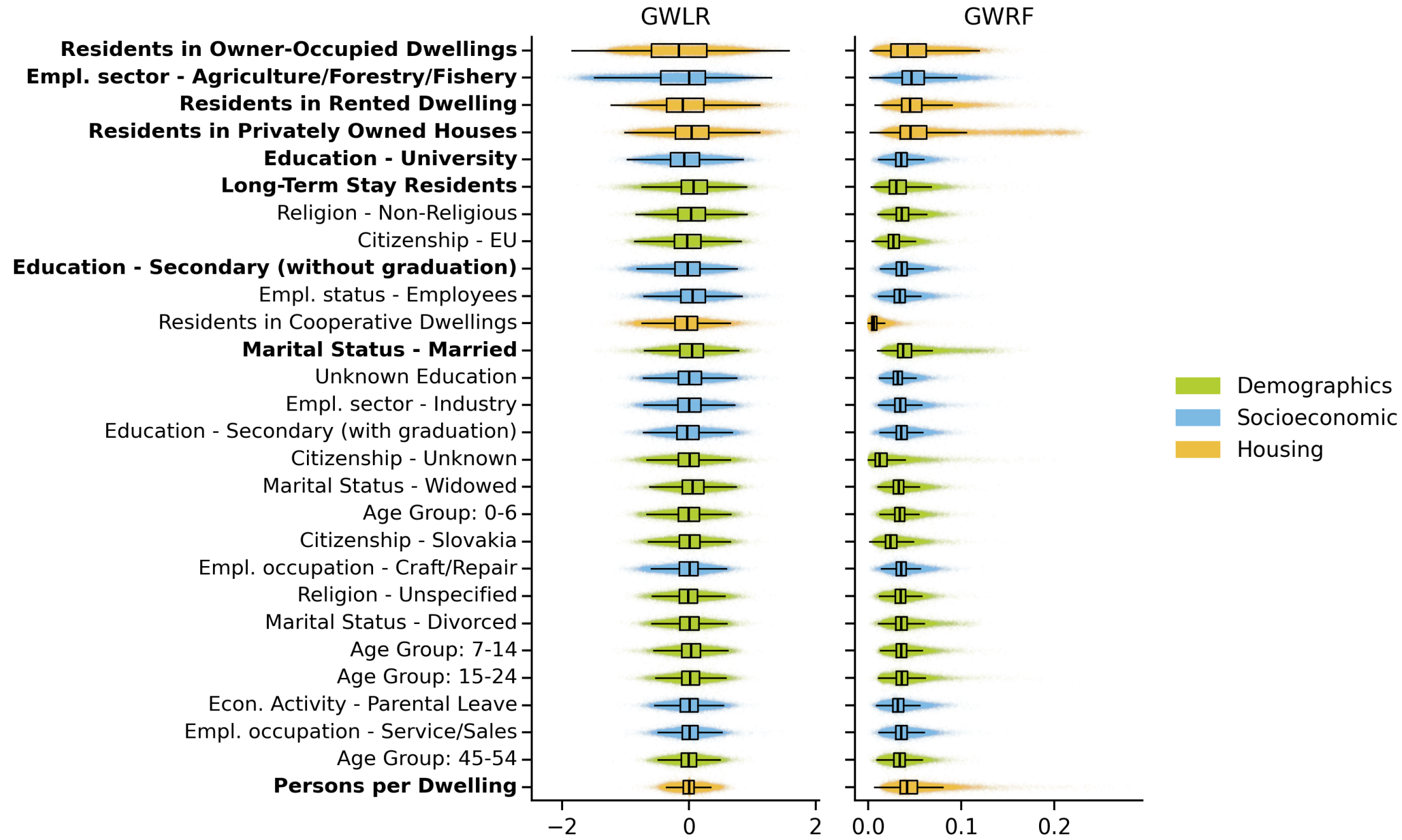

*Figure 5: Distribution of the local regression coefficients showing variability of their effect depending on location ordered by their magnitude to indicate overall influence and color-coded by their domain. Variables in bold indicate those with high spatial variability, where the standard deviation of their local coefficients falls within the upper quartile for either model.*

The Figure 5 shows that housing variables are among the most influential variables in both modelling approaches. However, these variables also exhibit distributional variance, suggesting that their effects are highly dependent on spatial context.

This is further shown in Figure 6, which displays the magnitude of a subset of local regression coefficients and local feature importances across space, alongside the varying directions of the regression coefficients for two variables identified as among the most influential, within two distinct built form types.

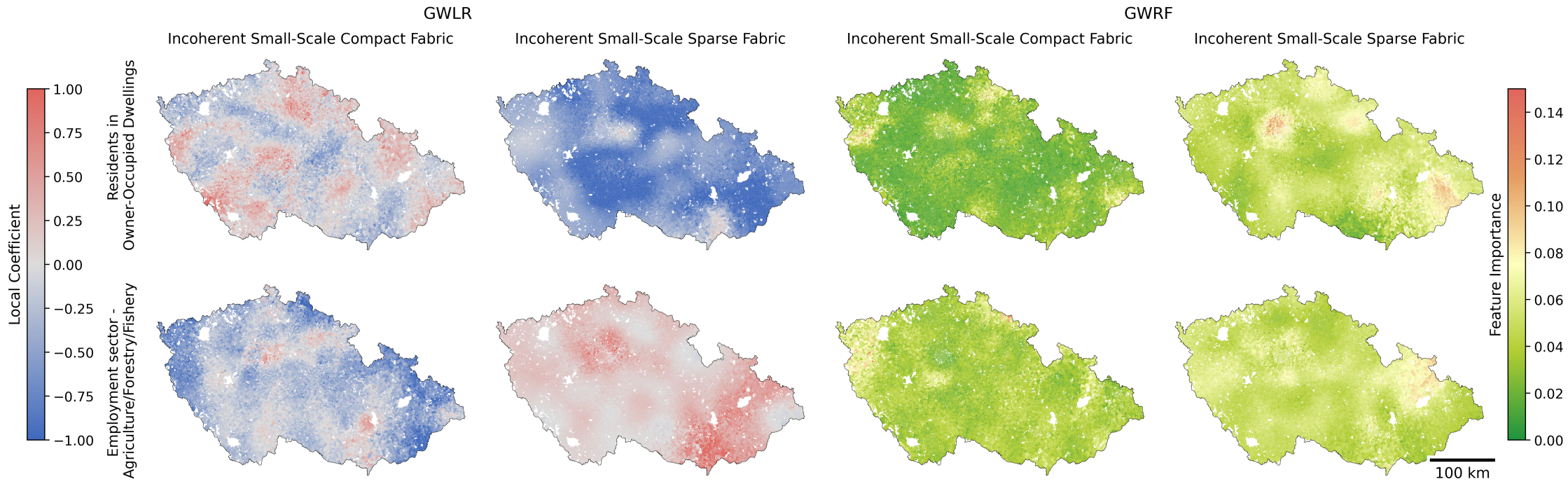

*Figure 6: Varying effects of the variables' local regression coefficients and local feature importances within certain built forms.*

Furthermore, we assessed which built form types are most sensitive to spatial context in the population–form associations by examining the distributions of the standard deviations of local regression coefficients and local feature importances across all variables for each built form type. The median of each box reflects the overall degree of spatial variability in the population–form associations for that built form type, while the width of each box reflects how uniformly this variability is distributed across variables. As shown in Figure 7, *Coherent Interconnected Fabric*, followed by *Coherent Dense Disjoint Fabric*, exhibit greater spatial variability in the effects of population characteristics, suggesting that the same variables operate differently across space within these forms. Moreover, the large interquartile range of standard deviations across variables within *Coherent Interconnected Fabric* suggests that this spatial instability is unevenly distributed — some variables shift substantially across space while others remain relatively stable. On the other hand, *Coherent Dense Adjacent Fabric* or *Incoherent Small-Scale Sparse Fabric* point to more spatially stable associations, with a narrow interquartile range indicating that most variables behave consistently across space within these forms.

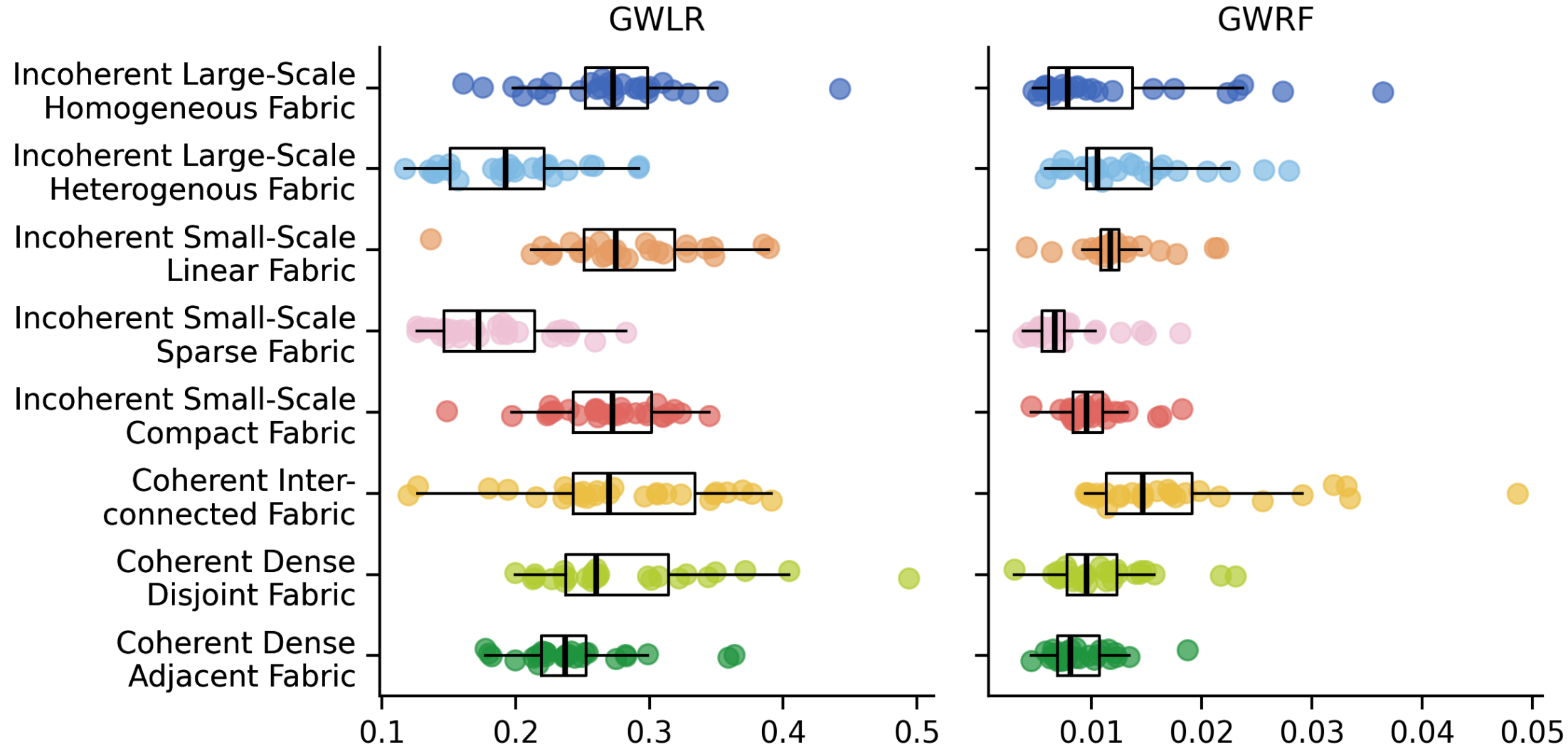

*Figure 7: Distribution of standard deviations of local regression coefficients and local feature importances across built-form types. The median reflects overall spatial variability in the population–form associations, while the width of each box reflects how uniformly this variability is distributed across variables.*

# 4 Discussion & Conclusion

First and foremost, the results indicate the necessity of modelling the relationship in a manner appropriate to its complex nature. The weak performance of global models points to the limitations of statistical approaches that assume spatial stationarity. By enforcing a single, global relationship, these models fail to capture the geographical heterogeneity rooted in the distinct historical development of Czechia's regions (Ouředníček, 2007; Sýkora & Bouzarovski, 2012; Temelová et al., 2011). On the other hand, the use of localised models significantly improves classification performance, providing empirical evidence of the inherently spatial nature of the relationship between population and built form. These results support the arguments of Edward Soja (1980) and David Harvey (1973), who argue that social processes are embedded in space and cannot be reduced to universal relationships that apply across different areas.

The comparative analysis of geographically weighted models shows that population characteristics exhibit mainly linear relationships with built form on the local level. As shown in the results, GWLR consistently outperforms or matches GWRF, emphasising the primary advantage of maintaining a linear approach – its clarity, as each coefficient represents a direct, quantifiable relationship.

The analysis of the strength of the associations underscores the importance of the physical layout of buildings, streets, and plots in reproducing socio-spatial inequalities. The results in Table 2 indicate that certain built form types might be more socially selective than others, attracting populations with specific profiles, whilst those with less pronounced characteristics may function as more socially fluid spaces. Altogether, this supports the literature, which argues that the physical environment is not socially neutral but actively contributes to the reproduction of social differences (Talen, 2008).

Whereas the associations are comparably strong across the study area for all built form types, they are not uniform within individual built forms. Sensitivity to local context varies considerably for different built form types, as shown in Figure 7. The *Coherent Interconnected Fabric*, typical for city centres, shows the highest sensitivity, suggesting that the associations are strongly shaped by local socio-demographic conditions. *Coherent Dense Adjacent Fabric*, typical for many suburban areas, on the other hand, shows lower spatial sensitivity, indicating that the association between population and these built forms is more uniform across space. The magnitude of the local feature metrics further reveals that indicators of housing tenure, education, and economic activity have the strongest overall influence in distinguishing between built form types (as shown in Figure 5). The distribution of these metrics' standard deviations confirms that the same socio-demographic factors can have different effects across local contexts.

The identification of housing tenure variables among the most influential indicators is consistent with the literature on housing inequalities in post-socialist countries, where housing tenure is closely connected to social inequalities (Lux & Sunega, 2025; Sunega & Lux, 2018). According to Lux & Sunega (2025), housing tenure is recognised as a key dimension of social differentiation that actively shapes residential choices and spatial patterns of inequality. In the Czech context, this reflects the long-term effects of large-scale privatisation, deregulation, and institutionalised support for homeownership (Lux & Sunega, 2010).

Despite these insights, the study faces several limitations. First, the analysis is affected by the spatial scale of the available population data. Although the Basic Settlement Units are the most granular spatial units, they remain coarse compared to building-level data. This level of aggregation assumes a certain degree of homogeneity within the population that may not exist, limiting the ability to capture subtle differences in the relationships between the population and built form. Second, some built form types are less frequently present at certain locations, leading to small sample sizes in geographically weighted models and limiting their effectiveness in estimating local relationships. Third, the population characteristics used are limited to those available in census data, excluding potentially relevant dimensions (e.g., income distribution, health conditions, ethnic background, wealth status, immigrant population) that could further explain socio-spatial differentiation.

While the analysis reveals which populations are most influential in distinguishing between built form types, it is difficult to derive specific socio-economic profiles associated with each type at each location. Future work could build on this research by explicitly linking built-form types to specific socio-demographic groups, thereby translating these associations into directly interpretable patterns of residential differentiation.

In conclusion, this study provides insight into the spatially conditioned dynamics between population structure and the built environment. We show that the population characteristics exhibit linear relationships with built form on the local level. We emphasise that spatial heterogeneity must be accounted for when assessing the associations between population and built form, and highlight the importance of localised modelling approaches to capture the complex relationships.

# Code and data availability

All components of the work rely on open source software and partially on open data, with the resulting code being openly available. Census data access is restricted and can be accessed from the Czech Statistical Office. Urban Taxonomy is available from doi.org/10.5281/zenodo.17076283. Code, with the specification of a reproducible environment, is available at https://github.com/uscuni/population-form-dynamics and archived at doi.org/10.5281/zenodo.18960572.

# Acknowledgments

The authors kindly acknowledge funding by the Charles University's Primus programme through the project "Influence of Socioeconomic and Cultural Factors on Urban Structure in Central Europe", project reference PRIMUS/24/SCI/023.

We would further like to thank to Lisa Winkler for support during initial phase of the project.